\documentstyle[preprint,prl,aps,epsf,tighten]{revtex}
\begin{document}
\title{Comment on ``Spin Transport properties of the quantum
one-dimensional non-linear sigma model''}
\author{Subir Sachdev$^{1}$ and Kedar Damle$^{2}$}
\address{$^1$Department of Physics, Yale University\\
P.O. Box 208120,
New Haven, CT 06520-8120, USA\\
$^2$Department of Electrical Engineering, Princeton University\\
Princeton, NJ 08544, USA\\}
\date{\today}
\maketitle
\begin{abstract}
In a recent preprint (cond-mat/9905415), Fujimoto
has used the Bethe ansatz to compute the finite temperature,
zero frequency Drude weight of spin transport in
the quantum ${\rm O}(3)$ non-linear sigma model in a magnetic field $H \neq 0$.
We show here that,
contrary to his claims, the results are in accord with
earlier semiclassical results (Sachdev and Damle, Phys. Rev. Lett.
{\bf 78}, 943 (1997)). We also comment on his $1/N$ expansion, and
show that it does not properly describe the long-time
correlations.
\end{abstract}
\pacs{PACS numbers:}

In a recent preprint~\cite{fuji}, Fujimoto has considered non-zero
temperature ($T$) transport in the one-dimensional quantum ${\rm O}(3)$
non-linear sigma
model. He considers the frequency ($\omega$) dependent
spin-conductivity, $\sigma (\omega)$, and tests for the
possibility that it
has a term of the form
\begin{equation}
\mbox{Re} \,\sigma ( \omega ) = K \delta (\omega) + \ldots.
\label{e1}
\end{equation}
In the presence of a non-zero magnetic field, $H \neq 0$, he uses
a Bethe ansatz computation to show in the low-temperature limit
that $K \sim \sqrt{T} e^{-(\Delta-H)/T}$,
where $\Delta$ is the magnitude of the $T=0$ energy gap.

Here we will show that, contrary to the claims of
Fujimoto~\cite{fuji}, this result is in precise accord with
earlier semiclassical results~\cite{sd}.
For a classical system, the dynamical spin conductivity is given in terms
of the the time ($t$) autocorrelation of
the total spin current $J(t)$ as
\begin{equation}
\sigma (\omega) = \frac{1}{TL} \int_0^{\infty} \langle J(t) J(0)
\rangle e^{i \omega t} dt,
\label{e1a}
\end{equation}
where $L$ is the size of the system, and, in the notation of
Ref~\onlinecite{sd}, the spin current is
\begin{equation}
J(t) = \sum_k m_k \frac{d x_k (t)}{dt},
\label{e2}
\end{equation}
where $m_k$ are the azimuthal spins of classical particles on
trajectories $x_k (t)$. Then the average over spins given in Eqn~3
of Ref~\onlinecite{sd} shows that
\begin{equation}
\langle J(t) J(0) \rangle = A_1 \sum_{k,\ell} \left \langle
\frac{d x_k (t)}{dt} \frac{d x_{\ell} (0)}{dt} \right\rangle
+ A_2 \sum_k \left \langle
\frac{d x_k (t)}{dt} \frac{d x_{k} (0)}{dt} \right\rangle.
\label{e3}
\end{equation}
We will now show that the first term proportional to $A_1$ above
contributes only to $K$; the second term proportional to $A_2$
yields only regular diffusive contributions to $\sigma (\omega)$,
and these latter terms were the focus of attention in Ref~\onlinecite{sd}.
The terms proportional to $A_1$ were also discussed in
Ref~\onlinecite{sd}, but Fujimoto appears to have overlooked them.
In the semiclassical model, the set of velocities at time $t$ is simply
a permutation of the velocities at $t=0$, and so in the first summation
in (\ref{e3}) we can relabel the
particles at time $t$ such that $dx_{k=\ell} (t)/dt = dx_{\ell} (0)/dt$.
Then the
average in the first term in (\ref{e3}) easily evaluates to an
average over a single Maxwell-Boltzmann distribution, and we get
\begin{equation}
\langle J(t) J(0) \rangle = A_1 \frac{L \rho c^2 T}{\Delta} + A_2 (
\ldots ),
\label{e4}
\end{equation}
where $c$ is the velocity of `light' in the sigma model, and $\rho$
is the total density of particles.
Combining (\ref{e1}), (\ref{e1a}) and (\ref{e4}), and using
expressions in Ref~\onlinecite{sd}, we have
\begin{equation}
K = \sqrt{ \frac{\pi T c^2}{2 \Delta}} e^{-(\Delta-H)/T}
\left(\frac{1 - 2 e^{-2H/T} + e^{-4H/T}}{1 + e^{-H/T} + e^{-2H/T}}\right).
\label{e5}
\end{equation}
This result is valid for
$H,T \ll \Delta$, but $H/T$ arbitrary. In the low $T$ limit
at fixed $H \neq 0$
($T \ll H \ll \Delta$), (\ref{e5}) agrees precisely with Fujimoto's
result for $K$.

It is interesting that $K$ vanishes as $H \rightarrow 0$ for
fixed $T \ll \Delta$, and
then the
conductivity only has the diffusive contribution proportional to
$A_2$~\cite{sd}.
Fujimoto has only quoted results in the low temperature
limit for fixed $H \neq 0$, and it would be
interesting to extend his computations to $H=0$ to access the complementary
regime discussed in Ref~\onlinecite{sd}.
Strictly speaking, a purely semiclassical method cannot rule
out the possibility that neglected quantum interference effects in special
integrable systems will lead to
a small non-zero $K$ at $H=0$, but we can expect that $K$
should at least be suppressed by factors of order (thermal
de Broglie wavelength)/(spacing between particles) from
its nominal $H \neq 0$ values.
Purely
diffusive transport is possible only at $H=0$, and more generally
in models with strict particle-hole symmetry~\cite{sd,ds,bs}.
It is interesting that a similar phenomenon has been noted in
the interacting electron models by Zotos {\em et al}~\cite{zotos},
who were able to prove ballistic transport only
in models without particle-hole symmetry.

Next, we comment on the $1/N$ expansion of transport properties.
Any kind of bare $1/N$ expansion~\cite{fuji}, or even the solution
of a $1/N$-derived quantum Boltzmann equation~\cite{qhe}, is doomed
to failure at low $T$ due to non-perturbative effects special to one spatial dimension.
Transport involves collisions of particles, and at low $T$
two-particle collisions dominate. The exact $S$-matrix~\cite{zama} for such
collisions is known at general $N$ --- it is ${\cal S}_{m_1' m_2'}^{m_1 m_2}
(\theta)$
where $\theta$ is a rapidity difference, and particles with spins
$m_1$, $m_2$ scatter into particles with spins $m_1'$, $ m_2'$.
Now for large $N$, at fixed $\theta$, we have
\begin{equation}
{\cal S}_{m_1' m_2'}^{m_1 m_2} (\theta) = \delta_{m_1 m_1'} \delta_{m_2
m_2'} + {\cal O}(1/N)
\end{equation}
which corresponds to ballistic {\em transmission of spin}, along with a
small amount of scattering at order $1/N$. However at low $T$,
small rapidities dominate, and we should really take the limit
$\theta \rightarrow 0 $ at fixed $N$. In this case we find, for
any fixed $N$
\begin{equation}
\lim_{\theta \rightarrow 0}
{\cal S}_{m_1' m_2'}^{m_1 m_2} (\theta) =(-1) \delta_{m_1 m_2'} \delta_{m_2
m_1'}.
\end{equation}
This corresponds to {\em total reflection of spin}, and was the key
effect behind the diffusive behavior discovered in
Ref~\onlinecite{sd}. This effect will not be captured at any
finite order in the $1/N$ expansion; this makes all conclusions drawn
from the $1/N$ expansion in Ref~\onlinecite{fuji} unreliable.

\end{document}